\documentclass[epjST]{svjour}
\usepackage{graphicx}
\usepackage{amsmath}
\usepackage{mathtools}
\usepackage{amssymb}
\usepackage{bm}
\usepackage{pifont}
\usepackage{color}
\usepackage[FIGTOPCAP]{subfigure}
\usepackage[latin1]{inputenc}

\begin{document}
%
%
%
%

\title{Discussion on Ohta et al., ``Traveling bands in self-propelled soft particles''}

\author{Thomas Ihle\inst{1}\inst{2}\fnmsep\thanks{\email{thomas.ihle@ndsu.edu}} \and
Yen-Liang Chou\inst{2}}

\institute{Department of Physics, North Dakota State University, Fargo, ND 58108-6050, USA \and
Max-Planck-Institute for the Physics of Complex Systems, N{\"o}thnitzer Stra{\ss}e 38, 01187 Dresden, Germany}

\abstract{
A discussion on the contribution of Ohta and Yamanaka \cite{ohta_ST} in this special issue, supplemented
by new agent-based simulations of band collisions within the standard Vicsek-model.
}

\maketitle

\section{Motivation}
\label{intro}

In contribution \cite{ohta_ST} (see also \cite{yamanaka_14}), the authors introduce a model of self-propelled deformable particles with a soft-core and
alignment interactions.
Performing agent-based numerical simulations,
they study the formation of different kinds of traveling bands and, in particular, they investigate
the behavior of these solitary bands in head-on collisions.
This is an interesting study; it extends our own work on band collisions \cite{ihle_13} in the simplistic Vicsek-model (VM) 
\cite{vicsek_95}
to more realistic systems of particles with (soft) excluded volume interactions and non-spherical shape. 

In the conclusion of contribution \cite{ohta_ST} the authors contrast their results 
with our previous results \cite{ihle_13}
and report opposite behavior. Specifically, they observe that bands of different size become of comparable size in subsequent collisions
whereas we predicted that the initial height difference of the bands amplifies in collisions leading to the scenario
of ``larger eats smaller".

In the first part of this comment, we would like to point out that the different behavior is not a contradiction
because we believe the models used in Ref. \cite{ohta_ST} and previous papers \cite{yamanaka_14,ohta_09,itino_11} are in  a 
different category than
the regular Vicsek-model with polar alignment.
In the second part, we would like to rule out possible errors on our side and report on corresponding 
agent-based simulations by the Vicsek-model.
In Ref. \cite{ihle_13} we showed quantitative agreement between kinetic theory and agent-based simulations for one
single, stationary solitary wave in the limit of large mean-free path and 
studied head-on collisions by numerically integrating the kinetic equations.
However, until now, we have never compared these collision studies to agent-based simulations.
This missing link is now presented here. These new simulations confirm our previous observations: 
In the parameter region we explored, we never saw that the relative height difference between 
two waves of significantly different size 
decreases. Here, the height difference was always measured at sufficiently large separation of the two waves.

\section{Model analysis and comparison to Vicsek model}

In this model \cite{ohta_ST,ohta_09}, the shape of a particle with label $i$ is phenomenologically described by means of a deformation tensor 
$S_{\alpha\beta}^{(i)}$ that becomes relevant when particle-particle interactions occur.
This tensor encodes the degree of deformation $s^{(i)}$ and ${\bf n}^{(i)}$, the unit normal along the long axis of the deformed particle.
The evolution equations for the velocities and deformation tensors, Eqs. (2)-(4) in \cite{ohta_ST},
can be rewritten in terms of the particle speeds $v^{(i)}$, deformations $s^{(i)}$, flying direction $\Phi^{(i)}$ 
(the angle between the particle
velocity and the $x$-axis) and the angle $\theta^{(i)}$ of the unit normal 
of the particle with respect to the $x$-axis as given in Eqs. (4)-(8) of Ref. \cite{ohta_09}.

Analysing the stationary states of these coupled equations 
for a single particle 
at the parameter values used in this contribution, $a=1$, $b=0.5$, $\gamma_0=\kappa=1$, one finds 
that
(i) the particle undergoes stable ballistic motion (assuming zero noise $\eta$) with constant speed $v_0$ and elongates when flying; the larger $v_0$ 
the more elongated it becomes,
(ii), the flying direction is parallel/anti-parallel to ${\bf n}^{(i)}$, that is $\phi^{(i)}=\theta^{(i)}\pm \pi$, and the particle travels into the direction of its largest semi-axis.
The force and noise terms are in the equation for the particle velocities. 
Thus, if a particle is slowed down by another one or by an obstacle, $s^{(i)}$ decreases and the particle becomes more spherical. 
Depending on the impact parameter of such an interaction, both the particle orientation $\theta^{(i)}$ and the flying direction 
$\phi^{(i)}$ might change.

Thus, in the model of Ohta {\em et al.} the traveling state of a single particle is characterized by four degrees of freedom, 
that are nontrivially and nonlinearly coupled, whereas the VM has only one parameter per particle: the traveling direction.

Given the many couplings present in the model of Ohta {\em at al.}, the system has the potential to exhibit a much 
richer dynamics than simple Vicsek-models.
Its particles might
do funny things under stress, for example when they run into a wall, have a collision with an incoming dense front of particles and so on.
While the authors wrote quite a number of papers on their model, \cite{ohta_09,itino_11,yamanaka_14}, 
performed a linear stability analysis and studied the transition to 
rotational motion, we were not able to find publications that systematically study collisions of just two particles or the 
interaction of one particle
with a wall as a function
of impact angle and model parameters.
This could be very helpful for understanding the solitary wave collisions on a microscopic level.
For example, one could ask what the maximum shape change of a particle is, can it become strongly 
squeezed in a collision, e.g. changes from 
prolate to oblate shape with negative $s^{(i)}$, 
and then escapes sideways? Can it switch temporarily to rotational motion when hit by others?
What are the relaxation times to recover from collision-induced deformation compared to the time interval until the next collision?
Recently, it also has been pointed out that short-range repulsion can induce a density-dependent particle speed and this coupling 
between speed and density can lead to a zoology of complex patterns \cite{peruani_klauss_11,farrell_12,barre_14}.
We think it would be worthwhile to discuss and compare the current model with respect to these developments.

Apart from the additional degrees of freedom, there is a more fundamental difference to the Vicsek-model (VM):
the symmetry of the alignment interactions. Recent classifications of active matter \cite{peruani_11,putzig_14,chate_08} 
for self-propelled particles without volume exclusion
distinguish between (A) 
nematic objects with nematic interactions, (B) polar particles with polar alignment and, (C) polar particles with nematic alignment.
The VM is in class (B), the nematic VM of Ref. \cite{peruani_08,ginelli_10} is in category (C). 

At first sight, the current model seems to be in class (C) since the alignment rule, Eq. (11) of \cite{ohta_ST}, has nematic symmetry, i.e. stays invariant 
if the angles $\theta$ of the involved particles are changed to $\theta\pm \pi$.
However, if this were the whole truth, the observed polarly ordered moving bands in Fig. 6 would be 
in contradiction with the results for the nematic VM \cite{ginelli_10} where stationary bands do not move and only show nematic order
(see Fig. 2(c) of \cite{ginelli_10}).
The patterns obtained by Ohta {\em et al.} are actually closer 
to the ones obtained for polar particles with polar alignment and a speed that is density dependent.
Thus, if one accepts the simple classification of dry active matter into three classes, the current model of deformable particles 
looks like a mixed case of (B) and (C) for the following reasons:

The collision rule of the nematic VM, Eq. (1) of \cite{ginelli_10}, describes an idealized case of fast orientational relaxation.
If one assumes the noise to be zero for the simplicity of this argument, perfect nematic 
alignment is already achieved in a single collision step.
This does not have to be the case in a more realistic interaction. 
From Eqs. (4-8) of \cite{ohta_09} one can read off finite relaxation times
for the director relaxation, deformation and so on.
If one assumes a grazing binary collision with $\theta^{(i)}\approx -\theta^{(j)}$,
the relative particle speed is of order $2v_0$ and the particles only have a contact time of order 
$\tau_C\sim \sigma/v_0$ to attempt alignment. Here, $\sigma$ is the effective radius of the deformable particle.
If $\tau_C$ is smaller than the relaxation time for alignment, the alignment will be incomplete. 
In contrast, if the particles have a parallel grazing collision, with $\theta^{(i)}\approx \theta^{(j)}$
they will stay together much longer and achieve much better alignment.
Thus, there is nematic alignment but it is biased towards parallel configurations.
As a result, the interactions can be seen as a perfect nematic alignment plus a small polar alignment.
We expect the relative importance of the polar component to be small for long rods, that is, particles with large aspect ratio.

Even though the particles in \cite{ohta_ST} are defined as points, the Gaussian soft-core potential, Eq. (9) together
with the asymmetric factor $Q_{ij}$ Eqs. (8), makes the particles interact like soft ellipsoids.
Using Eqs. (7-11) of \cite{ohta_ST} 
with interaction strength $Q=50$ and the single-particle deformation $s_0=bv_0^2/\kappa=0.4$, 
$v_0^2=\gamma_0/(1+ab/(2\kappa))=0.8$, it is possible to estimate the aspect ratio of these ellipsoids as between 2 and 3.
This is quite small, the associated ellipsoidal shape of a particle is not too far from a sphere, and
the polar component is likely be nonnegligible. 
This could contribute to the existence of polar bands as opposed to the immobile nematic band of Ref. \cite{ginelli_10}.
One could also speculate that a polar bias breaks the symmetry between parallel/antiparallel alignment and could be crucial 
even if very small. In any case, the qualitative difference between band collisions in the VM and the model of Ohta {\em et al.} raises
interesting questions about universality in active matter. 

Another difference between Vicsek-like models such as the one studied in \cite{peruani_08,ginelli_10} 
and more realistic models such as the current model \cite{ohta_ST} is that in the former,
particles can experience ``frontal collisions'' without much impact on their trajectories because
of the absence of volume exclusion.
Once there is at least a soft volume exclusion, nematic clusters where 50\% of the particles come from the right and 50\% from
the left cannot exist. 
This would also support the occurence of polar instead of purely nematic bands in Ref. \cite{ohta_ST}. 

In addition, the parameter $\kappa=1$ that controls the relaxation of a deformation
is not large and, hence, the particles are presumably very soft.
Because of the mixture of polar and nematic effects mentioned above and the softness of the particles, it is not totally surprising
that the bands behave qualitatively different than in the regular Vicsek model.
It would be interesting to isolate the possible reasons. For example, by increasing $\kappa$, 
ballistic motion should still be stable but
the particles are less soft and one could check the influence of softness on solitary wave collisions. 

\begin{figure}
\centering
\includegraphics[width=10.0cm,clip]{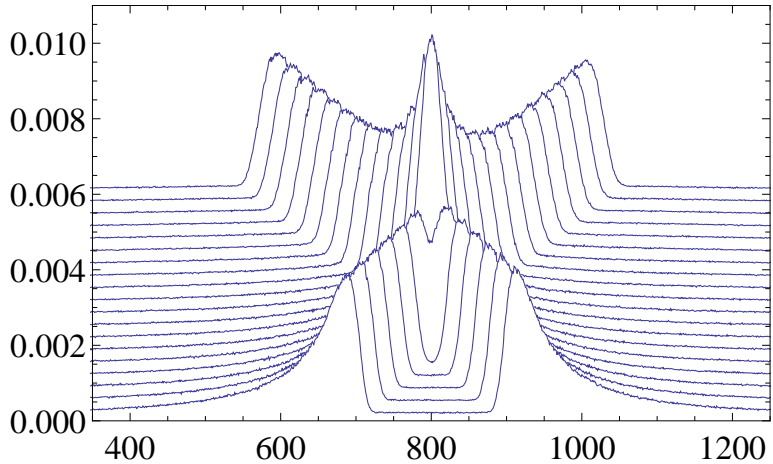}
\caption{
Density snapshots for the head-on collision of two soliton-like waves obtained in agent-based simulations.
The sequence starts with two well separated peaks close to the x-axis running
towards collision with their steep fronts facing each other.
At the latest time, the peaks are separated again after a successful ``tunneling'' through each other and now run towards
the edges of the box.
There is periodic boundary conditions in both $x$ and $y$ direction.
Parameters (defined in Ref. \cite{ihle_ST}): $M=\pi R^2\rho_0=0.0393$, $\tau v=2$, noise $\eta=0.46$, $L_x=1600$, $L_y=400$, particle number $N=8000$.
}
\label{FIG1}
\end{figure}

\begin{figure}
\centering
\includegraphics[width=10.0cm,clip]{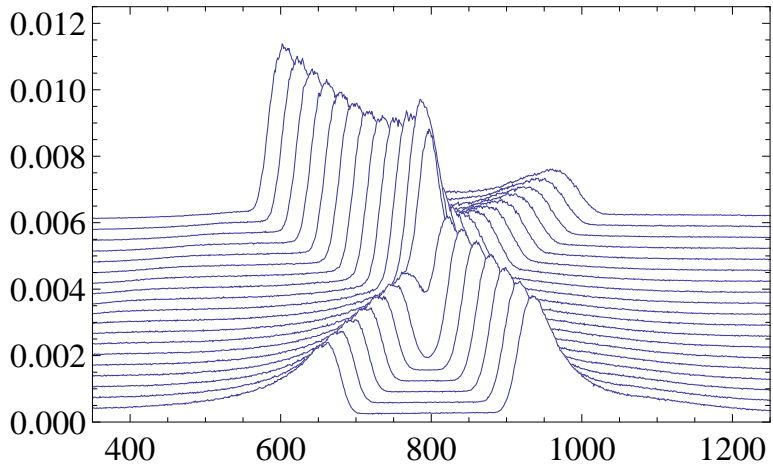}
\caption{Several snapshots of the density field of two colliding waves with significant height difference as a function of time 
for agent-based simulations of the VM.
Parameters: $M=0.0393$, $\tau v=2$, $\eta=0.40$, $L_x=1800$, $L_y=400$, $N=9000$.
}
\label{FIG2}
\end{figure}

The  existence of a short range repulsion with effective radius $\sigma=1$ 
in the model of Ohta {\em et al.} \cite{ohta_ST}
could be responsible for the
flat region at the top of the density wave in Fig. 6 (b). For the VM, where particles have zero volume, the wave top is very spiky \cite{ihle_13}, see also Figs. \ref{FIG1} and \ref{FIG2}.
\begin{figure}
\centering
\includegraphics[width=10.0cm,clip]{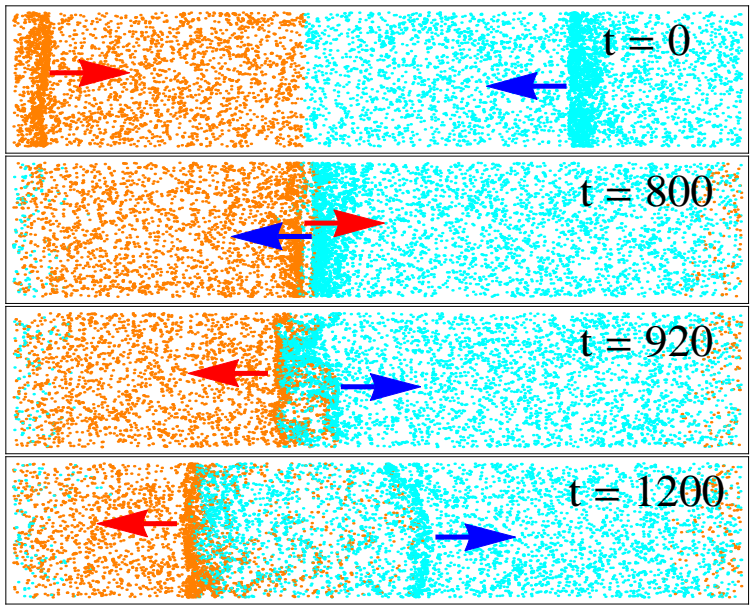}
\caption{Snapshots of two colliding waves at different for large density $M=0.7854$.
Only 20 \% of all particles are shown.
Parameters: $\tau v=0.5$, $\eta=1.6$, $L_x=1000$, $L_y=200$, $N=50000$.
}
\label{FIG3}
\end{figure}

\section{Agent-based simulations}

Here we repeat the ``soliton collision test'' as performed in \cite{ihle_13} but now using agent-based simulations of the standard 
Vicsek-model instead of numerically solving the kinetic equation of the one-particle distribution function.
We prepared stationary waves in two different systems with 
sizes $L_x^{(1)}$ and $L_x^{(2)}$, particle numbers $N_1$ and $N_2$, and ensured the waves run in opposite directions.
After the waves became stationary, the two boxes were ``glued'' together leading to a longer
system with $L_x=L_x^{(1)}+L_x^{(2)}$. 
A series of snapshots of the time evolution of the density, averaged
both over the $y$-direction and ensemble-averaged \cite{foot2}, is shown in
Fig. \ref{FIG1} for very small initial height difference of the waves.
At the earliest time, one sees two peaks running towards each other. Eventually,
they start to overlap and form a large single peak.
A while later, the two peaks reemerge with almost undisturbed shape like a conventional soliton.
Watching the time evolution through repeated collisions reveals that if the waves have a tiny height difference initially,
this difference is amplified in every encounter, as predicted by kinetic theory \cite{ihle_13}.
This increase of the height difference in a head-on collision becomes very clear in Fig. \ref{FIG2},
where the two waves have quite different sizes already at the beginning.
We found that this scenario is quite robust, even at larger densities, Fig. \ref{FIG3}, and different noise $\eta$.
However, at lower noise $\eta$, we observed a few cases, where both waves become smaller and slower in every collision, even though
the relative height difference did not decrease. We interpret this behavior as the possibly discontinuous 
phase transition from an inhomogeneous ordered phase with solitons to the homogeneously ordered phase that is expected at small noise.
\begin{figure}
\centering
\includegraphics[width=10.0cm,clip]{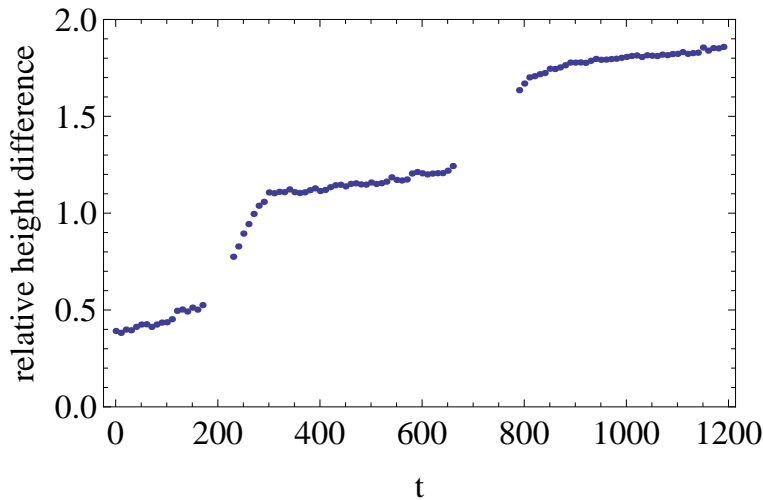}
\caption{Relative height difference $2|h_1-h_2|/(h_1+h_2)$ versus time for the runs shown in Fig. \ref{FIG2}.
The maximum densities of the two waves are given by $h_1$ and $h_2$, respectively.}
\label{FIG_HEIGHT}
\end{figure}
Fig. \ref{FIG_HEIGHT} shows the relative height difference for parameters corresponding to Fig. \ref{FIG2}.
When the solitary waves collide, the height difference jumps up abruptly. 
No data points were taken right after the overlap of the wave peaks to allow the waves to ``disentangle'' and relax to two separate waves again.

By using large lateral lengths up to $L_y=2000$ (not shown) 
and aspect ratios $L_y/L_x$ up to $1.5$ we made sure to allow for the possible formation of waves
going into the 
orthogonal $y$-direction after collision, something which is not possible by construction in the quasi-one dimensional 
runs of Ref. \cite{ihle_13}. While we observed lateral fluctuations of the wave fronts, see Fig. \ref{FIG3} bottom, that are 
usually straight in systems with small $L_y$, we never saw waves that switch their propagation direction like Fig. 7(e) in Ohta {\em et al.} \cite{ohta_ST}.
In large systems with $L_y/L_x=1.5$ we find that the time to recover to a straight wave front after collision is larger
than the time $L_x/v_w$ for the next collision that happens because of the periodic boundary condition over the shorter $L_x$ direction. 
Here, $v_w$ is the wave velocity. 
This velocity is larger than the speed of sound $v_S$ in the disordered phase, $v_S=v_0/\sqrt{2}$ \cite{ihle_13}.
As a consequence of the incomplete wave recovery, wave fronts show large fluctuations including partial break ups, until only one wave survives that does not undergo further collisions
and thus has enough time to become straight.
Even though we were not able to observe the scenarios reported by Ohta and Yamanaka, we cannot completely rule out this switch of
wave direction as well as the
possibility that 
bands of different size become of comparable size in subsequent collisions.
This is because the relevant parameter space in density, noise and mean free path is quite large, 
and we mostly focused on parameters close to the ones used in \cite{ihle_13}. 

To understand how solitons survive collisions, we ``painted'' the particles coming from the two different initial boxes
in different colors, orange and cyan. For simplicity of the argument, we assume here that both waves have the same height (although this is 
{\em not} the case in Fig. \ref{FIG3}, see \cite{foot1}
).
At first sight, it seems as if both particle groups are reflected from the
line where the wave fronts meet. However, looking very closely at the collision of the wave fronts in Fig. \ref{FIG3}
and corresponding videos, we suggest a different mechanism:
When the two wave fronts reach each other, small well aligned groups of particles penetrate the opposite front by a distance
of the order of the mean free path $\lambda=\tau v_0$, where $\tau$ is the time step of the Vicsek model and $v_0$ is the particle speed.
Since the density peak is very sharp, these groups are facing oppositely 
moving particles of the other color, that have a slightly smaller density and are thus slightly less aligned. Therefore, on average, these first penetrating groups manage to ``overpower''
the incoming particles in the tail of the opposite wave front and 
align them the other way \cite{foot1}. Now, after these groups have been reinforced by
particles of opposite color, they ``sweep up'' the rest of the incoming tail particles similar to a snowplow.
This explains why the wave fronts in Fig. \ref{FIG3} after head-on collision seem to consist of two layers. 
The leading front contains now the newly piled up
particles and in the tail one can see reminiscences of the original initiators of the change, which eventually fall behind and disappear from the main part of the wave.
It would be interesting to see the difference between this mechanism and what is going on in the model of \cite{ohta_ST}.


Support
from the National Science Foundation under grant No.
DMR-0706017 
is gratefully acknowledged.
We would like to thank Fernando Peruani for valuable discussions.


\begin{thebibliography}{99}
\bibitem{foot1}
Note, that the particle number per interaction circle of the VM is strongly fluctuating,
in fact, even stronger than in an ideal gas, see \cite{ihle_ST}. Without fluctuations, 
the above argument could not explain how a small wave can survive confrontation with a much bigger wave.
\bibitem{foot2}
In order to minimize artificial wave-broadening due to small speed differences of the waves, a specific type of ensemble average was 
used: The initial condition was fixed 1000 time steps before the two subsystems were ``glued'' together, 
and runs with different random number initializations were started from that position.
We verfied that snapshots from single realizations do not deviate too much from the ensemble average.
\end{thebibliography}
\end{document}